\documentclass[12pt]{article}
\usepackage{a4wide}
\usepackage{latexsym}
\usepackage{amsmath}
\usepackage{amsfonts}
\usepackage{amscd}
\usepackage{cite}

\usepackage{pslatex}
\usepackage{graphicx}
\usepackage[latin1]{inputenc}
\usepackage[T1]{fontenc}

\def\bq{\begin{eqnarray}}
\def\eq{\end{eqnarray}}

\def\eps{\varepsilon}
\def\v{\verb}

\begin{document}

\thispagestyle{empty}

\begin{flushright}
  MZ-TH/07-17 \\
\end{flushright}

\vspace{1.5cm}

\begin{center}
  {\Large\bf Resolution of singularities for multi-loop integrals\\
  }
  \vspace{1cm}
  {\large Christian Bogner ${}^{a,b}$ and Stefan Weinzierl ${}^{a}$\\
  \vspace{1cm}
      {\small ${}^{a}$ \em Institut f{\"u}r Physik, Universit{\"a}t Mainz,}\\
      {\small \em D - 55099 Mainz, Germany}\\
  \vspace{2mm}
      {\small ${}^{b}$ \em Department of Mathematical Sciences, University of Durham}\\
      {\small \em Durham DH1 3LE, United Kingdom}\\
  } 
\end{center}

\vspace{2cm}

\begin{abstract}\noindent
  {
   We report on a program for the numerical evaluation of divergent multi-loop integrals.
   The program is based on iterated sector decomposition.
   We improve the original algorithm of Binoth and Heinrich such that the program
   is guaranteed to terminate.
   The program can be used to compute numerically the Laurent expansion of divergent
   multi-loop integrals regulated by dimensional regularisation.
   The symbolic and the numerical steps of the algorithm are combined into one program.
   }
\end{abstract}

\vspace*{\fill}

\newpage 

{\bf\large PROGRAM SUMMARY}
\vspace{4mm}
\begin{sloppypar}
\noindent   {\em Title of program\/}: \v/sector_decomposition/ \\[2mm]
   {\em Version\/}: 1.0 \\[2mm]
   {\em Catalogue number\/}: \\[2mm]
   {\em Program obtained from\/}: {\tt http://wwwthep.physik.uni-mainz.de/\~{}stefanw/software.html} \\[2mm]
   {\em E-mail\/}: {\tt stefanw@thep.physik.uni-mainz.de}, {\tt bogner@thep.physik.uni-mainz.de}\\[2mm]
   {\em License\/}: GNU Public License \\[2mm]
   {\em Computers\/}: all \\[2mm]
   {\em Operating system\/}: Unix \\[2mm]
   {\em Program language\/}: {\tt C++     } \\[2mm]
   {\em Memory required to execute\/}: 
         Depending on the complexity of the problem. \\[2mm]
   {\em Other programs called\/}: GiNaC, available from {\tt http://www.ginac.de}, \\
         GNU scientific library, available from {\tt http://www.gnu.org/software/gsl}. \\[2mm]
   {\em External files needed\/}: none \\[2mm]
   {\em Nature of the physical problem\/}: 
         Computation of divergent multi-loop integrals. \\[2mm]
   {\em Method of solution\/}: 
         Sector decomposition.\\[2mm] 
   {\em Restrictions on complexity of the problem\/}: 
         Only limited by the available memory and CPU time. \\[2mm]
   {\em Typical running time\/}:
         Depending on the complexity of the problem.
\end{sloppypar}

\newpage

\section{Introduction}
\label{sec:intro}

The calculation of higher-order corrections in perturbation theory in particle physics
relies to a large extent on our abilities to compute Feynman loop integrals.
Any calculation of loop integrals is complicated by the fact, that these integrals may contain
divergences, ultraviolet or infrared in nature.
Dimensional regularisation is usually employed to regulate the divergences 
and one seeks the Laurent expansion of Feynman integrals in $D=4-2\eps$ dimensions
in the small parameter $\eps$.

Automated numerical algorithms, which permit the computation of the coefficients of the Laurent series
for certain phase space points are of great use as an independent check of analytical calculations.
There are several methods available for this task.
All methods have to face the challenge to disentangle overlapping singularities.
In this paper we focus on sector decomposition\cite{Hepp:1966eg,Roth:1996pd,Binoth:2000ps,Binoth:2003ak}.
Other possibilities include a method based on 
difference equations\cite{Laporta:2000dc,Laporta:2001dd,Laporta:2001rc,Laporta:2002pg}
or the evaluation of Mellin-Barnes integrals
\cite{Czakon:2005rk,Anastasiou:2005cb}.

Sector decomposition
is a convenient tool to disentangle
overlapping singularities. 
Binoth and Heinrich \cite{Binoth:2000ps,Binoth:2003ak}
gave an algorithm for the automated computation of the Laurent series
of multi-loop integrals.
The method has recently also been applied to 
phase space integrals\cite{Gehrmann-DeRidder:2003bm,Binoth:2004jv,Heinrich:2006sw} 
and complete loop amplitudes\cite{Lazopoulos:2007ix,Anastasiou:2007qb}.

However, the original formulation of Binoth and Heinrich has a drawback: The disentanglement of
overlapping singularities is achieved through recursive sector decomposition. 
If this process terminates, the algorithm gives the correct answer.
But there is no guarantee that the algorithm stops after
a finite number of steps.
In fact, one can find relatively simple counter-examples, where the original algorithm leads to an
infinite loop.
In this paper we improve the method and present an algorithm, which is guaranteed to terminate.
This is possible, since we can relate our problem to a well-known problem in algebraic geometry:
The resolution of singularities of an algebraic variety over a field of characteristic zero
by a sequence of blow-ups.
In view of \cite{Bloch:2006} a connection between Feynman integrals and algebraic geometry
comes to no surprise.
The mathematical problem can be solved, which was first shown by Hironaka \cite{Hironaka:1964}.
Although the proof by Hironaka is non-constructive and therefore cannot be used as an algorithm, 
in subsequent years constructive methods have been 
developed \cite{Spivakovsky:1983,Villamayor:1989,Villamayor:1992,Bierstone:1997,Encinas:1998,Encinas:2000,Encinas:2002,Bravo:2003,Hauser:2003,Zeillinger:PhD,Zeillinger:2006}.
Let us mention for completeness that the mathematical problem is more general,
the disentanglement of singularities of multi-loop integrals corresponds to the sub-case, where
the algebraic variety is generated through a single polynomial.
We have implemented three different strategies for the resolution of the singularities.
All of them are guaranteed to terminate and yield the correct answer.
They differ in the required CPU-time.

The algorithm for sector decomposition can be divided into two stages.
The first stage involves
symbolic computer algebra, while the second stage is based on a numerical integration.
Partially due to this ``double nature'' of the algorithm, 
no public program is available up to now.
Given the importance and the popularity of the algorithm, this gap
should be closed.
With this article we provide such a program.

A user-friendly implementation of the algorithm would require that the different aspects of the method,
e.g.~the symbolic and the numerical computations, are treated within the same frame-work.
The library GiNaC\cite{Bauer:2000cp} allows the symbolic manipulation of expressions within the programming
language C++.
It offers the tools to combine symbolic and numerical computations in one program.
Furthermore there exist specialised libraries for an efficient numerical Monte Carlo integration\cite{GSL,Kawabata:1995th,Hahn:2004fe}. 
Our implementation is therefore in C++, using the libraries GiNaC for the symbolic part and
the GNU Scientific Library for the numerical part.
All stages of the algorithm are implemented within one program.

This article is organised as follows:
In section~\ref{sect:feynman_integrals} we review basic facts about multi-loop Feynman integrals.
Section~\ref{sect:algorithm} gives a brief description of the algorithm for sector decomposition.
In section~\ref{sect:strategies} we discuss in detail strategies, which guarantee a termination
of the iterated sector decomposition.
Section~\ref{sect:program} gives an overview of the design of the program
while section~\ref{sect:howto}
is of a more practical nature and describes how to install and use the program.
This section provides an example and compares the efficiency of the different strategies.
Finally, a summary is provided in section~\ref{sect:summary}.
In an appendix we describe technical details of the implementation
and provide a proof for one of the strategies (the proofs for the two other strategies
can be found in the literature).


\section{Feynman integrals}
\label{sect:feynman_integrals}

In this section we briefly recall some basic facts about multi-loop Feynman integrals.
A generic scalar $l$-loop integral $I_G$ 
in $D$ dimensions with $n$ propagators corresponding to a graph $G$
is given by
\bq
\label{eq0}
I_G  & = &
 \int \prod\limits_{r=1}^{l} \frac{d^Dk_r}{i\pi^{\frac{D}{2}}}\;
 \prod\limits_{j=1}^{n} \frac{1}{(-q_j^2+m_j^2)^{\nu_j}}.
\eq
The momenta $q_j$ of 
the propagators are linear combinations of the external momenta and the loop
momenta.
After Feynman parameterisation one arrives at
\bq
\label{eq1}
I_G  & = &
 \frac{\Gamma(\nu-lD/2)}{\prod\limits_{j=1}^{n}\Gamma(\nu_j)}
 \int\limits_{x_j \ge 0}  d^nx \;
 \delta(1-\sum_{i=1}^n x_i)\,
 \left( \prod\limits_{j=1}^n x_j^{\nu_j-1} \right)
 \frac{{\mathcal U}^{\nu-(l+1) D/2}}{{\mathcal F}^{\nu-l D/2}},
\eq
with $\nu=\nu_1+...+\nu_n$.
The functions ${\mathcal U}$ and $\mathcal F$ depend on the Feynman parameters.
If one expresses
\bq
 \sum\limits_{j=1}^{n} x_{j} (-q_j^2+m_j^2)
 & = & 
 - \sum\limits_{r=1}^{l} \sum\limits_{s=1}^{l} k_r M_{rs} k_s + \sum\limits_{r=1}^{l} 2 k_r \cdot Q_r - J,
\eq
where $M$ is a $l \times l$ matrix with scalar entries and $Q$ is a $l$-vector
with four-vectors as entries,
one obtains
\bq
 {\mathcal U} = \mbox{det}(M),
 & &
 {\mathcal F} = \mbox{det}(M) \left( - J + Q M^{-1} Q \right).
\eq
Alternatively,
the functions ${\mathcal U}$ and ${\mathcal F}$ can be derived 
from the topology of the corresponding Feynman graph $G$.
Cutting $l$ lines of a given connected $l$-loop graph such that it becomes a connected
tree graph $T$ defines a chord ${\mathcal C}(T,G)$ as being the set of lines 
not belonging to this tree. The Feynman parameters associated with each chord 
define a monomial of degree $l$. The set of all such trees (or 1-trees) 
is denoted by ${\mathcal T}_1$.  The 1-trees $T \in {\mathcal T}_1$ define 
${\mathcal U}$ as being the sum over all monomials corresponding 
to the chords ${\mathcal C}(T,G)$.
Cutting one more line of a 1-tree leads to two disconnected trees $(T_1,T_2)$, or a 2-tree.
${\mathcal T}_2$ is the set of all such  pairs.
The corresponding chords define  monomials of degree $l+1$. Each 2-tree of a graph
corresponds to a cut defined by cutting the lines which connected the two now disconnected trees
in the original graph. 
The square of the sum of momenta through the cut lines 
of one of the two disconnected trees $T_1$ or $T_2$
defines a Lorentz invariant
\bq
s_{T} & = & \left( \sum\limits_{j\in {\mathcal C}(T,G)} p_j \right)^2.
\eq   
The function ${\mathcal F}_0$ is the sum over all such monomials times 
minus the corresponding invariant. The function ${\mathcal F}$ is then given by ${\mathcal F}_0$ plus an additional piece
involving the internal masses $m_j$.
In summary, the functions ${\mathcal U}$ and ${\mathcal F}$ are obtained from the graph as follows:
\bq
\label{eq0def}	
 {\mathcal U} 
 & = & 
 \sum\limits_{T\in {\mathcal T}_1} \Bigl[\prod\limits_{j\in {\mathcal C}(T,G)}x_j\Bigr]\;,
 \nonumber\\
 {\mathcal F}_0 
 & = & 
 \sum\limits_{(T_1,T_2)\in {\mathcal T}_2}\;\Bigl[ \prod\limits_{j\in {\mathcal C}(T_1,G)} x_j \Bigr]\, (-s_{T_1})\;,
 \nonumber\\
 {\mathcal F} 
 & = &  
 {\mathcal F}_0 + {\mathcal U} \sum\limits_{j=1}^{n} x_j m_j^2\;.
\eq
In general, ${\mathcal U}$ is a positive semi-definite function. 
Its vanishing is related to the  UV sub-divergences of the graph. 
Overall UV divergences, if present,
will always be contained in the  prefactor $\Gamma(\nu-l D/2)$. 
In the Euclidean region, ${\mathcal F}$ is also a positive semi-definite function 
of the Feynman parameters $x_j$.  
The vanishing of ${\mathcal F}$ is related to infrared divergences.
Note that this is only a necessary but not sufficient condition for the occurrence
of an infrared singularity.
If an infrared singularity occurs or not depends in addition on the external kinematics.


\section{A review of the algorithm for sector decomposition}
\label{sect:algorithm}

In this section we briefly describe the original algorithm of Binoth and Heinrich for iterated
sector decomposition.
A detailed discussion on strategies for choosing the sub-sectors is deferred to section~\ref{sect:strategies}.
The program calculates the Laurent expansion in $\eps$ of integrals of the form
\bq
\label{basic_integral}
 \int\limits_{x_j \ge 0} d^nx \;\delta(1-\sum_{i=1}^n x_i)
 \left( \prod\limits_{i=1}^n x_i^{a_i+\eps b_i} \right)
 \prod\limits_{j=1}^r \left[ P_j(x) \right]^{c_j+\eps d_j}.
\eq
The integration is over the standard simplex.
The $a$'s, $b$'s, $c$'s and $d$'s are integers.
The $P$'s are polynomials in the variables $x_1$, ..., $x_n$.
The polynomials are required to be non-zero
inside the integration region, but
may vanish on the boundaries of the integration region.
Note that the program allows a product of several polynomials and that it is not required
that the polynomials are homogeneous.
\\
\\
Step 0: We first convert all polynomials to homogeneous polynomials.
Due to the presence of the delta-function we have
\bq
 1 & = & x_1 + x_2 + ... + x_n.
\eq
For each polynomial $P_j$ we determine the highest degree $h_j$ and multiply all terms with a lower
degree by an appropriate power of $x_1 + x_2 + ... + x_n$.
As a result, the polynomial $P_j$ is then homogeneous of degree $h_j$.
\\
\\
Step 1: We decompose the integral into $n$ primary sectors as in
\bq
 \int\limits_{x_j \ge 0} d^nx & = &
 \sum\limits_{l=1}^n \int\limits_{x_j \ge 0} d^nx 
     \prod\limits_{i=1, i\neq l}^n \theta(x_l \ge x_i).
\eq
In the $l$-th primary sector we make the substitution
\bq
 x_j & = & x_l x_j' \;\;\;\mbox{for} \; j \neq l.
\eq
As each polynomial $P_j$ is now homogeneous of degree $h_j$ we arrive at
\bq
\lefteqn{
 \int\limits_{x_j \ge 0} d^nx \;\delta(1-\sum_{i=1}^n x_i)
 \left( \prod\limits_{i=1, i\neq l}^n \theta(x_l \ge x_i) \right)
 \left( \prod\limits_{i=1}^n x_i^{a_i+\eps b_i} \right)
 \prod\limits_{j=1}^r \left[ P_j(x) \right]^{c_j+\eps d_j}
= } & & \nonumber \\
 & & 
 \int\limits_0^1 
 \left( \prod\limits_{i=1, i\neq l}^n dx_i \; x_i^{a_i+\eps b_i} \right)
 \left( 1 + \sum\limits_{j=1, j\neq l}^n x_j \right)^c
 \prod\limits_{j=1}^r \left[ P_j(x_1,...,x_{j-1},1,x_{j+1},...,x_n) \right]^{c_j+\eps d_j},
\eq
where
\bq
 c & = & -n - \sum\limits_{i=1}^n \left( a_i+\eps b_i \right) - \sum\limits_{j=1}^r h_j \left( c_j+\eps d_j\right).
\eq
Each primary sector is now a $(n-1)$-dimensional integral over the unit hyper-cube.
Note that in the general case this decomposition introduces an additional polynomial factor
\bq
 \left( 1 + \sum\limits_{j=1, j\neq l}^n x_j \right)^c.
\eq
However for Feynman integrals of the form as in eq.~(\ref{eq1}) we always have $c=0$ and this
factor is absent. In any case, this factor is just an additional polynomial.
In general, we therefore deal with integrals of the form
\bq
\label{primary_integral}
 \int\limits_{0}^{1} d^{n}x
  \prod\limits_{i=1}^{n}x_{i}^{a_{i}+\epsilon b_{i}}
  \prod\limits_{j=1}^{r} \left[P_{j}(x)\right]^{c_{j}+\epsilon d_{j}}.
\eq
Step 2: We decompose the primary sectors iteratively into sub-sectors until each of the polynomials 
is of the form
\bq
\label{monomialised}
 P & = & x_1^{m_1} ... x_n^{m_n} \left( 1 + P'(x) \right),
\eq
where $P'(x)$ is a polynomial in the variables $x_j$.
If $P$ is of the form~(\ref{monomialised}), we say that $P$ is monomialised.
In this case the monomial prefactor $x_1^{m_1} ... x_n^{m_n}$ can be factored out
and the remainder contains a constant term.
To convert $P$ into the form~(\ref{monomialised}) we choose a subset
$S=\left\{ \alpha_{1},\,...,\, \alpha_{k}\right\} \subseteq \left\{ 1, \,...\, n \right\}$
according to one of the strategies outlined in section~\ref{sect:strategies}.
We decompose the $k$-dimensional hypercube into $k$ sub-sectors according to
\bq
 \int\limits_{0}^{1} d^{n}x & = & 
 \sum\limits_{l=1}^{k} 
 \int\limits_{0}^{1} d^{n}x
   \prod\limits_{i=1, i\neq l}^{k}
   \theta\left(x_{\alpha_{l}}\geq x_{\alpha_{i}}\right).
\eq
In the $l$-th sub-sector we make for each element of $S$ the
substitution
\bq
\label{substitution}
x_{\alpha_{i}} & = & x_{\alpha_{l}} x_{\alpha_{i}}' \;\;\;\mbox{for}\; i\neq l.
\eq
This yields a Jacobian factor $x_{\alpha_{l}}^{k-1}$. 
This procedure is iterated, until all polynomials are of the form~(\ref{monomialised}).
The strategies discussed in section~\ref{sect:strategies} guarantee that the recursion
terminates.
At the end all polynomials contain a constant term.
Each sub-sector integral is of the form as in eq.~(\ref{primary_integral}),
where every $P_{j}$ is now different from zero in the whole integration
domain. Hence the singular behaviour of the integral depends on the
$a_{i}$ and $b_{i}$, the $a_{i}$ being integers.
\\
\\
Step 3: For every $x_{j}$ with $a_{j}<0$ we have to perform a Taylor
expansion around $x_{j}=0$ in order to extract the possible $\epsilon$-poles.
If we consider for the moment only one parameter $x_{j}$ we 
can write the
corresponding integral as 
\bq
 \int\limits_{0}^{1} dx_{j} \; x_{j}^{a_{j}+b_{j}\eps} \mathcal{I}(x_{j})
 = 
 \int\limits_{0}^{1} dx_{j} \; x_{j}^{a_{j}+b_{j}\eps}
   \left(\sum\limits_{p=0}^{\left|a_{j}\right|-1} \frac{x_{j}^{p}}{p!} \mathcal{I}^{(p)} 
         + \mathcal{I}^{(R)}(x_j)
   \right)
\eq
where we defined 
$\mathcal{I}^{(p)} = \left. \partial/\partial x_{j}^{p} \mathcal{I}(x_{j})\right|_{x_{j}=0}$.
The remainder term
\bq
 \mathcal{I}^{(R)}(x_j) & = & 
   \mathcal{I}(x_{j}) - \sum\limits_{p=0}^{\left|a_{j}\right|-1} \frac{x_{j}^{p}}{p!} \mathcal{I}^{(p)}
\eq
does not lead to $\eps$-poles in the $x_{j}$-integration.
The integration in the pole part can be carried out analytically:
\bq
 \int\limits_{0}^{1} dx_{j} \; x_{j}^{a_{j}+b_{j}\eps}
   \; \frac{x_{j}^{p}}{p!} \mathcal{I}^{(p)} 
  & = &
   \frac{1}{a_{j}+b_{j}\eps+p+1} \frac{\mathcal{I}^{(p)}}{p!}.
\eq
This procedure is repeated for all variables $x_j$ for which $a_j<0$.
\\
\\
Step 4: All remaining integrals are now by construction finite.
We can now expand all expressions in a Laurent series in $\eps$
\bq
 \sum\limits_{i=A}^{B}C_{i}\epsilon^{i}+O\left(\epsilon^{B}\right)
\eq
and truncate to the desired order.
\\
\\
Step 5: It remains to compute the coefficients $C_{i}$.
These coefficients contain finite integrals, which we evaluate numerically
by Monte Carlo integration.


\section{Strategies for choosing the sub-sectors}
\label{sect:strategies}

In this section we discuss strategies 
for choosing the subset $S$ which defines
the sub-sectors, such that the iterated 
sector decomposition terminates.
Binoth and Heinrich proposed
to determine a minimal subset $S=\left\{ \alpha_{1},\,...,\, \alpha_{k}\right\} $
such that at least one polynomial $P_j$ vanishes for $x_{\alpha_{1}} = ... = x_{\alpha_{k}} = 0$.
By a minimal set we mean a set which does not contain a proper subset having this property.
If $S$ contains only one element, $S=\left\{ \alpha\right\}$, then the corresponding Feynman parameter
factorises from $P_j$.
A relative simple example shows, that this procedure may lead to an infinite loop:
If one considers the polynomial
\bq
P\left(x_{1},\, x_{2},\, x_{3}\right)=x_{1}x_{3}^{2}+x_{2}^{2}+x_{2}x_{3},
\eq
then the subset $S=\left\{ 1,\, 2\right\} $ is an allowed choice,
as $P\left(x_{1}=0,\, x_{2}=0,\, x_3\right)=0$ and $S$ is minimal. 
In the first sector the substitution (\ref{substitution}) reads 
$x_{1}=x_{1}^{\prime},\, x_{2}=x_{1}^{\prime}x_{2}^{\prime},\, x_{3}=x_{3}^{\prime}$.
It yields 
\bq
P\left(x_{1},\, x_{2},\, x_{3}\right) & = &
 x_{1}^{\prime}x_{3}^{\prime2}+x_{1}^{\prime2}x_{2}^{\prime2}+x_{1}^{\prime}x_{2}^{\prime}x_{3}^{\prime}
 =
 x_{1}^{\prime}\left(x_{3}^{\prime2}+x_{1}^{\prime}x_{2}^{\prime2}+x_{2}^{\prime}x_{3}^{\prime}\right)
 =
 x_{1}^{\prime}P\left(x_{1}^{\prime},\, x_{3}^{\prime},\, x_{2}^{\prime}\right).
\eq
The original polynomial has been reproduced, which leads to an infinite recursion.
The problem of a non-terminating iteration has already been noted in ref.~\cite{Binoth:2003ak}.
To avoid this situation we need a strategy for choosing $S$, for which we can show that
the recursion always terminates.
This is a highly non-trivial problem. 
It is closely related to the resolution of singularities 
of an algebraic variety over a field of characteristic zero.
Fortunately, mathematicians have found a solution for the latter problem
and we can adapt the mathematical solutions to our problem.
We will not go into any details of algebraic geometry.
Instead we present in section~\ref{sect:polyhedra_game} the polyhedra game,
introduced by Hironaka to illustrate the problem of resolution of singularities.
The polyhedra game can be stated with little mathematics.
Any solution to the polyhedra game will correspond to a strategy for choosing the subsets $S$.
In sections~\ref{sect:strategy_A} to \ref{sect:strategy_C} 
we present three winning strategies for the polyhedra game.
All three strategies ensure that the iterated sector decomposition terminates
and lead to the correct result.
However, the number of generated sub-sectors (and therefore the efficiency of the method)
will vary among the different strategies.
Common to all strategies is a sequence of positive numbers associated to the polynomials.
All strategies enforce this sequence to decrease with each step in the iteration with respect
to lexicographical ordering.
As the sequence cannot decrease forever, the algorithm is guaranteed to terminate.
The actual construction of this sequence will differ for different strategies.

The lexicographical ordering is defined as follows:
A sequence of real numbers $(a_1,...,a_r)$ is less than a sequence $(b_1,...,b_s)$ if and only if there
exists a $k \in {\mathbb N}$ such that
$a_j = b_j$ for all $j < k$ and $a_k < b_k$, where we use the convention that $a_j=0$ for $j>r$.

\subsection{Hironaka's polyhedra game}
\label{sect:polyhedra_game}

Hironaka's polyhedra game is played by two players, A and B. They are
given a finite set $M$ of points $m=\left(m_{1},\,...,\,m_{n}\right)$
in $\mathbb{N}_{+}^{n}$, the first quadrant of $\mathbb{N}^{n}$.
We denote by $\Delta \subset\mathbb{R}_{+}^{n}$ the positive convex hull of the set $M$.
It is given by the convex hull of the set
\bq
\bigcup\limits_{m\in M}\left(m+\mathbb{R}_{+}^{n}\right).
\eq
The two players compete in the following game:
\begin{enumerate}
\item Player A chooses a non-empty subset $S\subseteq\left\{ 1,\,...,\, n\right\}$.
\item Player B chooses one element $i$ out of this subset $S$. 
\end{enumerate}
Then, according
to the choices of the players, the components of all $\left(m_{1},\,...,\,m_{n}\right)\in M$
are replaced by new points $\left(m_{1}^{\prime},\,...,\,m_{n}^{\prime}\right)$,
given by:
\bq
\label{update_polyhedron}
m_{j}^{\prime} & = & m_{j}, \;\;\; \textrm{if }j\neq i, \nonumber \\
m_{i}^{\prime} & = & \sum_{j\in S} m_{j}-c,
\eq
where for the moment we set $c=1$. 
This defines the set $M^\prime$.
One then sets $M=M^\prime$ and goes back to step 1.
Player A wins the game if, after a finite number of moves, 
the polyhedron $\Delta$ is of the form
\bq
\label{termination}
 \Delta & = & m+\mathbb{R}_{+}^{n},
\eq
i.e. generated by one point.
If this never occurs, player $B$ has won.
The challenge of the polyhedra game is to show that player $A$ always has
a winning strategy, no matter how player $B$ chooses his moves.

Let us discuss the relation of Hironaka's polyhedra game to the sector decomposition of multi-loop
integrals.
Without loss of generality we can assume that we have just one polynomial $P$
in eq.~(\ref{basic_integral}).
(If there are several polynomials, we obtain a single polynomial by multiplying them together.
As only the zero-sets of the polynomials are relevant, the exponents can be neglected.)
The polynomial $P$ has the form
\bq
 P & = & 
  \sum\limits_{i=1}^p c_i x_1^{m_1^{(i)}} x_2^{m_2^{(i)}} ... x_n^{m_n^{(i)}}.
\eq
The $n$-tuple $m^{(i)}=\left(m^{(i)}_{1},\,...,\,m^{(i)}_{n}\right)$ defines a point in
$\mathbb{N}_{+}^{n}$ and 
$M=\left\{m^{(1)},\,...\,m^{(p)} \right\}$ is the set of all such points.
Substituting the
parameters $x_{j}$ according to equation~(\ref{substitution}) and factoring out
a term $x_{i}^c$ yields the same polynomial as replacing the powers
$m_{j}$ according to equation~(\ref{update_polyhedron}). 
In this sense, one iteration of the sector decomposition corresponds to one move in Hironaka's game. 
Reducing $P$ to the form~(\ref{monomialised})
is equivalent to achieving~(\ref{termination}) in the polyhedra game.
Finding a strategy which guarantees termination of the iterated sector decomposition 
corresponds to a winning strategy for player $A$ in the polyhedra game.
Note that we really need a strategy that guarantees player A's victory for every
choice player B can take, because the sector decomposition has to be carried out
in every appearing sector. In other words, we sample over all possible decisions of B.

\subsection{Strategy A}
\label{sect:strategy_A}

Zeillinger's strategy \cite{Zeillinger:PhD,Zeillinger:2006} for player $A$  is conceptionally the simplest.
A drawback of this strategy is that it leads to a large number of sub-sectors and is therefore quite
inefficient.
Denote by $M_C$ the set of corners of $\Delta$, $M_{C}\subseteq M$.
Define $W$ to be the set of vectors connecting
these corners:
\bq
W & = & \left\{ v=m^{(i)}-m^{(j)}:\,m^{(i)},\,m^{(j)} \in M_{C}\right\}.
\eq
We define the length $L(v)$ of the vector $v\in W$ by 
\bq
L(v) & = & \max_{1\leq i\leq n}v_{i}-\min_{1\leq i\leq n}v_{i}.
\eq
We define the multiplicity $N(v)$ of the vector $v\in W$ by the number
of its components equal to its minimal or maximal
component:
\bq
N(v) & = & \#\left\{ j\,:\, 1\leq j\leq n,\, 
                            v_j = \min_{1\leq i\leq n}\, v_{i} \;\;\mbox{or}\;\; 
                            v_j=\max_{1\leq i\leq n}\, v_{i}\right\}.
\eq
From the set $W$ we single out one vector $w$ 
for which the sequence of the two numbers 
$L\left(w\right)$ and $N\left(w\right)$
is minimal with respect to lexicographical ordering:
\bq
\left(L\left(w\right),\, N\left(w\right)\right)
 & \leq_{\textrm{lex}} &
 \left(L(v),\, N(v)\right), \;\;\;\forall v\in W.
\eq
Player A chooses then $S=\left\{ k,\, l\right\}$, where $k$ and $l$ are defined by
\bq
 w_k = \min_{1\leq i\leq n} w_i 
 \;\;\;\textrm{and}\;\;\; 
 w_l = \max_{1\leq i\leq n} w_{i}.
\eq
This choice reduces the sequence of the three numbers
\bq
b & = & 
 \left\{ \begin{array}{l}
  \left(0,\,0,\,0\right)
  \\
  \left(\# M_{C},\, L\left(v_{M}\right),\, N\left(v_{M}\right)\right)
 \end{array}
 \right.
 \begin{array}{l}
  \textrm{if }\Delta\;\textrm{is monomial,}\\
  \textrm{otherwise, }
 \end{array}
\eq
with respect to lexicographical ordering.
$\# M_{C}$ denotes the number of corners of $\Delta$.
A proof that this strategy is a solution to Hironaka's polyhedra game can be found 
in \cite{Zeillinger:PhD,Zeillinger:2006}.

\subsection{Strategy B}
\label{sect:strategy_B}

Spivakovsky's strategy \cite{Spivakovsky:1983} was the first solution to Hironaka's polyhedra game.
To state the algorithm, we need a few auxiliary definitions:
We define $\omega\left(\Delta\right)\in\mathbb{R}_{+}^{n}$ as 
the vector given by the minima of the individual coordinates of elements in $\Delta$: 
\bq
\omega_{i} & = & \textrm{min}\left\{ \nu_{i}\mid\nu\in\Delta\right\} ,\; i=1,\,...,\, n.
\eq
Furthermore we write 
$\tilde{\Delta}=\Delta-\omega\left(\Delta\right)$
and 
$\tilde{\nu}_{i}=\nu_{i}-\omega_{i}$.
For a subset $\Gamma\subseteq\left\{ 1,\,...,\, n\right\} $ we define
\bq
d_{\Gamma}\left(\Delta\right) & = &
 \textrm{min}\left\{ \sum_{j\in\Gamma}\nu_{j}\mid\nu\in\Delta\right\} 
 \;\;\;\textrm{and}\;\;\; 
 d\left(\Delta\right)=d_{\left\{ 1,\,...,\, n\right\} }\left(\Delta\right).
\eq
We then define a sequence of sets
\bq
\label{spivakovsky_set_sequence}
 \left( I_0, \Delta_0, I_1, \Delta_1, ..., I_r, \Delta_r \right)
\eq
starting from
\bq
 I_0 = \left\{ 1,\,...,\, n\right\},
 & &
 \Delta_0 = \Delta.
\eq
For each $\Delta_k$ we define a set 
$H_k$ by
\bq
H_k & = & 
 \left\{ j\in I_k 
         \mid \; \exists \; \nu\in\Delta_k 
              \textrm{ such that }\sum_{i \in I_k} \nu_{i}=d\left(\Delta_k\right)
              \textrm{ and }\tilde{\nu}_{j}\neq0\right\}.
\eq
$I_{k+1}$ is given by
\bq
 I_{k+1} & = & I_{k}\backslash H_k.
\eq
In order to define $\Delta_{k+1}$ we first define 
for the two complementary subsets $H_k$ and $I_{k+1}$ of $I_k$ the set
\bq
M_{H_k} & = &
 \left\{ \nu\in\mathbb{R}_{+}^{I_k} 
         \mid
         \sum_{j\in H_k}\nu_{j}<1\right\}
\eq
and the projection
\bq
 P_{H_k} & : & M_{H_k} \longrightarrow \mathbb{R}_{+}^{I_{k+1}},
 \nonumber \\
 & &
 P_{H_k}\left(\alpha,\,\beta\right)=\frac{\alpha}{1-\left|\beta\right|},
 \;\;\;
 \alpha \in \mathbb{R}_{+}^{I_{k+1}},
 \;\;\;
 \beta \in \mathbb{R}_{+}^{H_{k}},
 \;\;\;
 \left| \beta \right| = \sum\limits_{j\in H_k} \beta_j.
\eq
Then $\Delta_{k+1}$ is given by
\bq
\label{def_projected_set}
\Delta_{k+1} & = & 
 P_{H_k}
 \left(M_{H_k}
 \cap
  \left(\frac{\tilde{\Delta}_{k}}{d\left(\tilde{\Delta}_{k}\right)}\cup\Delta_{k}\right)
 \right),
\eq
where 
$\tilde{\Delta}_{k} = \Delta_k - \omega\left(\Delta_k\right)$.
The sequence in eq.~(\ref{spivakovsky_set_sequence}) stops if either 
$d\left(\tilde{\Delta}_{r}\right)=0$
or $\Delta_{r}=\emptyset$.
Based on the sequence in eq.~(\ref{spivakovsky_set_sequence}) player A chooses now the
set $S$ as follows:
\begin{enumerate}

\item If $\Delta_{r}=\emptyset$, player A chooses $S=\left\{ 1,\,...,\, n\right\} \backslash I_{r}$.

\item If $\Delta_{r}\neq\emptyset$, player A first chooses a minimal subset $\Gamma_{r}\subseteq I_{r}$,
such that $\sum_{j\in \Gamma_{r}}\nu_{j}\geq1$ for all $\nu\in\Delta_{r}$ and sets
$S=\left(\left\{ 1,\,...,\, n\right\} \backslash I_{r}\right)\cup \Gamma_{r}$.
\end{enumerate}
To each stage of the game (i.~e.~to each $\Delta$), we can associate
a sequence of $2r+2$ numbers
\bq
\delta\left(\Delta\right) & = &
 \left( d\left(\tilde{\Delta}\right),\,\# I_{1},\, d\left(\tilde{\Delta}_{1}\right),
        \,...,\,
        \# I_{r},\, d\left(\tilde{\Delta}_{r}\right),\, d\left(\Delta_r\right)\right),
\eq
adopting the conventions $\tilde{\emptyset}=\emptyset$ and $d\left(\emptyset\right)=\infty$.
The above strategy forces $\delta\left(\Delta\right)$
to decrease with each move in the sense of a lexicographical ordering.
Further, it can be shown that $\delta\left(\Delta\right)$
cannot decrease forever. Hence player
A is guaranteed to win. 
The proof is given in \cite{Spivakovsky:1983}.

\subsection{Strategy C}
\label{sect:strategy_C}

In this section we present a strategy inspired by the proof of Encinas and Hauser \cite{Encinas:2002,Hauser:2003}.
We keep as much as possible the notation of the previous section.
Similar to the previous section we define a sequence
\bq
\label{hauser_set_sequence}
 \left( c_{-1}, I_0, \Delta_0, c_0, I_1, \Delta_1, ..., c_{r-1}, I_r, \Delta_r \right)
\eq
starting from
\bq
 c_{-1} = 0,
 \;\;\;
 I_0 = \left\{ 1,\,...,\, n\right\},
 \;\;\;
 \Delta_0 = \Delta.
\eq
Compared to eq.~(\ref{spivakovsky_set_sequence}) we added the numbers $c_{-1}, c_0, ..., c_{r-1}$ to the sequence.
Again we define $H_k$ by
\bq
H_k & = & 
 \left\{ j\in I_k
         \mid \; \exists \; \nu\in\Delta_k 
              \textrm{ such that }\sum_{i \in I_k} \nu_{i}=d\left(\Delta_k\right)
              \textrm{ and }\tilde{\nu}_{j}\neq0\right\}.
\eq
We set $c_k=d\left(\tilde{\Delta}_k\right)$ and define 
for a set $\tilde{\Delta}_k$ a companion set $\tilde{\Delta}^C_k$ by
\bq
\label{def_companion}
 \tilde{\Delta}^C_k & = &
 \left\{
 \begin{array}{ll}
   \tilde{\Delta}_k \cup 
     \left( \frac{c_k}{c_{k-1}-c_k} \omega\left(\Delta_k\right) 
            + \mathbb{R}_{+}^{I_{k}} \right)
        & \mbox{if} \; 0 < c_k < c_{k-1}, \\
   \tilde{\Delta}_k & \mbox{otherwise}. \\
 \end{array}
 \right.
\eq
Let $T_k \subseteq H_k$ be a subset of $H_k$.
The subset $T_k$ is chosen according to a rule $R$: $T_k=R(H_k)$.
The only requirement on the rule $R$ is, that is has to be deterministic:
If
\bq
\label{choice_T_k}
 T=R(H)\;\;\mbox{and}\;\;T'=R(H')\;\;\mbox{and}\;\;H=H'\;\;\Rightarrow\;\;T=T'.
\eq
We define
\bq
 \Delta_{T_k} & = & 
   c_k \; P_{T_k} \left( M_{T_k} \cap \frac{\tilde{\Delta}^C_k}{c_k} \right).
\eq
We then set
\bq
 I_{k+1} & = & I_k \backslash T_k, 
  \nonumber \\
 \Delta_{k+1} & = & \Delta_{T_k}.
\eq
The sequence in eq.~(\ref{hauser_set_sequence}) stops if either 
$d\left(\tilde{\Delta}_{r}\right)=0$
or $\Delta_{r}=\emptyset$.
Based on the sequence in eq.~(\ref{spivakovsky_set_sequence}) player A chooses now the
set $S$ as follows:
\begin{enumerate}

\item If $\Delta_{r}=\emptyset$, player A chooses $S=\left\{ 1,\,...,\, n\right\} \backslash I_{r}$.

\item If $\Delta_{r}\neq\emptyset$, player A first chooses a minimal subset $\Gamma_{r}\subseteq I_{r}$,
such that $\sum_{j\in \Gamma_{r}}\nu_{j}\geq c_{r-1}$ for all $\nu\in\Delta_{r}$ and sets
$S=\left(\left\{ 1,\,...,\, n\right\} \backslash I_{r}\right)\cup \Gamma_{r}$.
\end{enumerate}
This strategy reduces the sequence
\bq
\label{sequence_Hauser}
 i(\Delta) & = &
 \left( d\left(\tilde{\Delta}_0\right), \left(\#I_0-\#H_0\right), 
        d\left(\tilde{\Delta}_{1}\right), 
        ...,
        \left(\#I_{r-1}-\#H_{r-1}\right),
        d\left(\tilde{\Delta}_{r}\right),
        d\left(\Delta_r\right)
 \right)
\eq
with respect to lexicographical ordering.
The strategy C is similar to strategy B, the major differences are in the choice
of the companion set eq.~(\ref{def_companion}) as compared to eq.~(\ref{def_projected_set})
and the freedom to choose a subset $T_k \subseteq H_K$ instead of $H_k$.
We provide a proof for this strategy in appendix \ref{appendix:proof}.


\section{A description of the program}
\label{sect:program}

The program is written in C++.
The main routine to compute an integral of the form as in eq.~(\ref{basic_integral}) 
is the function \v/do_sector_decomposition/. The arguments are as follows:
\begin{verbatim}
monte_carlo_result do_sector_decomposition(
                    const integration_data & global_data,
                    const integrand & integrand_in,
                    const monte_carlo_parameters & mc_parameters,
                    int verbose_level = 0);

\end{verbatim}
The input are three structures, \v/integration_data/, \v/integrand/,
\v/monte_carlo_parameters/, which
are described in detail below, and an optional parameter \v/verbose_level/.
With the help of the optional parameter \v/verbose_level/ one can choose the amount 
of information the program prints out during the run.
The function returns a structure \v/monte_carlo_result/, which again is described below.
The strategy for the iterated sector decomposition is selected by the global variable
\v/CHOICE_STRATEGY/. The keywords for the available strategies are
\begin{verbatim}
 STRATEGY_A, STRATEGY_B, STRATEGY_C, STRATEGY_X.
\end{verbatim}
The first three strategies are described in detail in the previous section and are guaranteed to terminate.
The last one is a heuristic strategy, for which we neither have a proof that it terminates,
nor do we know a counter-example which would lead to an infinite recursion.
This strategy chooses the smallest set $S$ such that the maximal power of a Feynman parameter 
can be factored out in each sub-sector.
It is included for the following reason: If this strategy terminates, the number of the generated sub-sectors
tends to be smaller than the corresponding numbers for the other strategies.
The default is
\begin{verbatim}
CHOICE_STRATEGY = STRATEGY_C;
\end{verbatim}
The class \v/integration_data/ contains the data which will not be modified
by the algorithm for the sector decomposition.
It has a constructor of the form
\begin{verbatim}
integration_data(const std::vector<GiNaC::ex> & list_feynman_parameter, 
                 GiNaC::ex epsilon, int order);
\end{verbatim}
where \v/list_feynman_parameter/ is a vector holding the symbols of the Feynman parameters,
\v/epsilon/ is a symbol corresponding to $\eps$ in eq.~(\ref{basic_integral})
and \v/order/ defines which term of the Laurent series in $\eps$ should be computed.
\\
\\
The integrand
\bq
 \left( \prod\limits_{i=1}^n x_i^{a_i+\eps b_i} \right)
 \prod\limits_{j=1}^r \left[ P_j(x) \right]^{c_j+\eps d_j}
\eq
is encoded in the class \v/integrand/. This class has a constructor
of the form
\begin{verbatim}
integrand(const std::vector<exponent> & nu,
          const std::vector<GiNaC::ex> & poly_list, 
          const std::vector<exponent> & c);
\end{verbatim}
where \v/nu/ is a vector of size $n$ holding the exponents $a_i+\eps b_i$.
\v/poly_list/ is a vector of size $r$, holding the polynomials $P_j$ in the Feynman parameters.
The corresponding exponents are stored in the vector \v/c/, again of size $r$.
As exponents are generally of the form $a+b\eps$, a special class is available for them:
\begin{verbatim}
exponent(int a, int b);
\end{verbatim}
In applications one encounters often integrands where the explicit powers of the 
Feynman parameters are missing. For integrands of the form
\bq
 \prod\limits_{j=1}^r \left[ P_j(x) \right]^{c_j+\eps d_j}
\eq
there is a simpler constructor of the form
\begin{verbatim}
integrand(size_t n, const std::vector<GiNaC::ex> & poly_list, 
          const std::vector<exponent> & c);
\end{verbatim}
Here $n$ is the number of Feynman parameters.
\\
\\
Parameters associated to the Monte Carlo integration are specified with the help of the class
\v/monte_carlo_parameters/.
This class is constructed as follows:
\begin{verbatim}
monte_carlo_parameters(size_t iterations_low, size_t iterations_high, 
                       size_t calls_low, size_t calls_high);
\end{verbatim}
The program uses the Vegas-algorithm\cite{Lepage:1978sw,Lepage:1980dq}
based on an adaptive grid. 
The program does first a Monte Carlo integration with \v/iterations_low/ iterations
with \v/calls_low/ function evaluations each. This phase is solely used to adapt the grid.
The numerical result of the Monte Carlo integration is then obtained from the
second stage with \v/iterations_high/ iterations of \v/calls_high/ function calls each.
\\
\\
The main function \v/do_sector_decomposition/ returns the numerical results
of the Monte Carlo integration in the class \v/monte_carlo_result/.
This class has the methods
\begin{verbatim}
  class monte_carlo_result {
    public :
      double get_mean(void) const;
      double get_error(void) const;
      double get_chi_squared(void) const;
  };
\end{verbatim}
which return the mean value of the Monte Carlo integration, the error estimate and 
the associated $\chi^2$.

\section{How to use the program}
\label{sect:howto}

In this section we give indications how to install and use the program library.
Compilation of the package will build a (shared) library. The user can then write his own
programs, using the functions provided by the library by linking his executables
against the library.

\subsection{Installation}

The program can be obtained from\\
\\
{\tt http://wwwthep.physik.uni-mainz.de/\~{}stefanw/software.html}\\
\\
After unpacking, the library for sector decomposition
is build by issuing
the commands
\begin{verbatim}
  ./configure 
  make
  make install
\end{verbatim}
There are various options which can be passed to the configure script,
an overview can be obtained with {\tt ./configure -{}-help}.\\
\\
After installation, the shell script \v/sector_decomposition-config/ can be used
to determine the compiler and linker command line options required
to compile and link a program with the library.
For example, \v/sector_decomposition-config --cppflags/ will give the path to the header files
of the library, whereas \v/sector_decomposition-config --libs/ prints out the flags
necessary to link a program against the library.

\subsection{Writing programs using the library}

The following test program computes the first terms of the Laurent series of the
massless double-box graph. The graph corresponds to the integral
\bq
 \int\limits_{x_j \ge 0}  d^7x \;
 \delta(1-\sum_{i=1}^7 x_i)\;
 {\mathcal U}^{1+3\eps} {\mathcal F}^{-3-2\eps},
\eq
with
\bq
 {\mathcal U} & = &
  (x_1+x_2+x_3)(x_5+x_6+x_7) + x_4 (x_1+x_2+x_3+x_5+x_6+x_7),
 \\
 {\mathcal F} & = &
  \left[x_2 x_3 (x_4+x_5+x_6+x_7)+x_5 x_6 (x_1+x_2+x_3+x_4)+x_2 x_4 x_6+x_3 x_4 x_5\right](-s)
    +x_1 x_4 x_7 (-t). \nonumber 
\eq
The integral is computed for the point $(s,t)=(-1,-1)$.
\begin{verbatim}
#include <iostream>
#include <vector>
#include <ginac/ginac.h>
#include "sector_decomposition/sector_decomposition.h"

int main()
{
  using namespace GiNaC;
  using namespace sector_decomposition;

  CHOICE_STRATEGY = STRATEGY_X;

  symbol eps("eps");
  symbol s("s"), t("t");
  symbol x1("x1"), x2("x2"), x3("x3"), x4("x4"), x5("x5"), x6("x6"), x7("x7");
  const ex x[] = { x1,x2,x3,x4,x5,x6,x7 };

  std::vector<ex> parameters(7);
  for (int i=0; i<7; i++) parameters[i]=x[i];

  ex U = (x1+x2+x3)*(x5+x6+x7) + x4*(x1+x2+x3+x5+x6+x7);
  ex F = (x2*x3*(x4+x5+x6+x7)+x5*x6*(x1+x2+x3+x4)+x2*x4*x6+x3*x4*x5)*(-s)
    +x1*x4*x7*(-t);

  F = F.subs(lst( s == -1 , t == -1 ));

  std::vector<ex> poly_list(2);
  poly_list[0] = U;
  poly_list[1] = F;

  std::vector<exponent> c(2);
  c[0] = exponent( 1, 3 );
  c[1] = exponent( -3, -2 );

  integrand my_integrand(7, poly_list, c);

  monte_carlo_parameters mc_parameters( 5, 15, 10000, 100000 );

  for (int order=-4; order<=0; order++)
    {
      integration_data global_data(parameters, eps, order);

      monte_carlo_result res = 
        do_sector_decomposition(global_data, my_integrand, mc_parameters);

      std::cout << "Order " << pow(eps,order) << ": " << res.get_mean() 
                << " +/- " << res.get_error() << std::endl;
    }

  return 0;
}
\end{verbatim}
Running the program will print out the following result:
\begin{verbatim}
Order eps^(-4): 2.00001 +/- 9.25208e-05
Order eps^(-3): -5.99992 +/- 0.000359897
Order eps^(-2): -4.91623 +/- 0.00157598
Order eps^(-1): 11.4958 +/- 0.00681643
Order 1: 13.8236 +/- 0.0207286
\end{verbatim}
The running time is about 40 minutes on a standard PC.

\subsection{Further examples and performance}

\begin{figure}[ht]
\begin{center}
\includegraphics[bb= 30 510 570 710,width=0.8\textwidth]{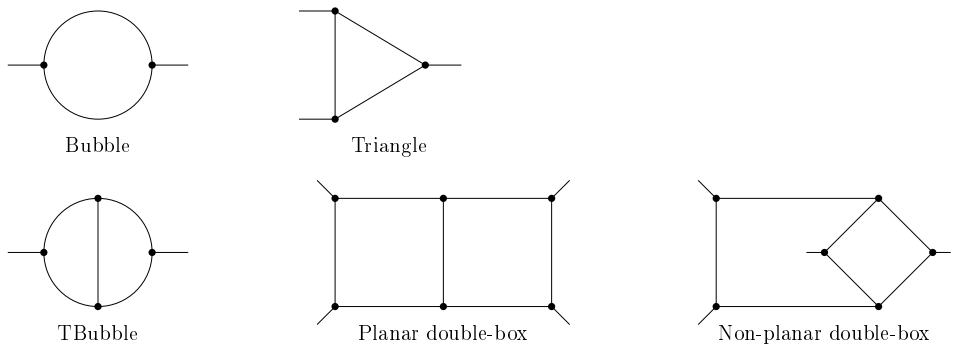}
\end{center}
\caption{Feynman integrals included as examples in the distribution.}
\label{fig1}
\end{figure}

The program comes with several examples: The one-loop two-point and three-point functions, as well as
the following two-loop functions: The two-point function, the planar double-box and the 
non-planar double box.
The corresponding Feynman diagrams for these examples are shown in fig.~\ref{fig1}.
For these examples the corresponding analytic results are known and can be found for the 
two-loop integrals in \cite{Bierenbaum:2003ud,Smirnov:1999gc,Tausk:1999vh}.
We have used these results to verify the correctness of our code.

The running time of the program is dominated by the numerical Monte-Carlo integration.
The running time for the Monte-Carlo integration depends on the complexity of the integrand,
which in turn is related to the number of generated sub-sectors.
The number of generated sub-sectors is therefore a measure for the efficiency of the algorithm.
Note that although different strategies lead to the same result for the Laurent expansion of
multi-loop integrals, the number of generated sub-sectors may differ among the 
various strategies.
In table~\ref{table1} we compare the number of generated sub-sectors for the different strategies and
for different loop integrals.
\begin{table}
\begin{center}
\begin{tabular}{l|rrrr}
 Integral & Strategy A & Strategy B & Strategy C & Strategy X \\
 \hline
 & & & & \\
 Bubble & 2 & 2 & 2 & 2 \\
 Triangle & 3 & 3 & 3 & 3 \\
 Tbubble & 58 & 48 & 48 & 48 \\
 Planar double-box & 755 & 586 & 586 & 293 \\
 Non-planar double-box & 1138 & 698 & 698 & 395 \\
\end{tabular}
\end{center}
\caption{The number of generated sub-sectors for different loop integrals and
different strategies.}
\label{table1}
\end{table}
We observe that strategy~A performs worse than strategies B or C.
Strategies A, B and C are guaranteed to terminate. Strategy X, for which we have no proof that the
singularities are resolved after a finite number of iterations, terminates for the examples above
and gives the lowest numbers for the generated sub-sectors.
For this reason it is included in the program.
A pragmatic approach would try strategy X first.
If the iterated sector decomposition terminates, one obtains the result efficiently.
If not, one uses as fall-back option one of the strategies $A$, $B$ or $C$.

Going to a higher number of loops and more propagators, the memory requirements become
an issue. However, the individual sub-sectors are independent and can be calculated once at a time.
This reduces significantly the memory requirements.
With this method we evaluated the massless on-shell triple box in about two days CPU time on a standard PC.

\subsection{Documentation}

The complete documentation of the program is inserted as comment lines in
the source code.
The documentation can be extracted from the sources
with the help of the documentation system ``doxygen'' \cite{doxygen}.
The program ``doxygen'' is freely available.
Issuing in the top-level build directory for the library the commands
\begin{verbatim}
  doxygen Doxyfile
\end{verbatim}
will create a directory ``reference'' with the documentation in html and latex format.


\section{Summary}
\label{sect:summary}

In this article we considered the numerical computation of the coefficients
of the Laurent expansion of a divergent multi-loop integral, regulated
by dimensional regularisation.
The method is based on iterative blow-ups, also known as iterative sector
decomposition.
The algorithms we employed ensure that this recursive procedure terminates.
We implemented the algorithms into a computer program, such that
the symbolic and the numerical steps of the algorithms are contained in one program.

\subsection*{Acknowledgements}

We would like to thank S. M\"uller-Stach and H. Spiesberger for
useful discussions.

\begin{appendix}

\section{Details on the implementation}

The numerical Monte Carlo integration is by far the most time-consuming part of the program
and efficiency of the program in this part is of particular importance.

The GiNaC-library provides a method for the numerical evaluation of a function, based on
arbitrary precision arithmetic.
For Monte Carlo integration, where a function needs to be evaluated
many times, this is quite slow and therefore inefficient.
It is also not needed, since statistical errors and not rounding
errors tend to dominate the error of the final result.
Therefore a different approach has been implemented for the numerical
Monte Carlo integration:
The function to be integrated is first written as C code to a file,
this file is then compiled with a standard C compiler and the resulting
executable is loaded dynamically (e.g. as a ``plug-in'') into the memory
space of the program and the Monte Carlo integration routine uses this
compiled C function for the evaluations.

The performance is further improved by optimising the C code of the function.
This is done by replacing subexpressions, which occur syntactically 
more than once by temporary variables. 
In addition, subexpressions with a large number of operands are split into smaller pieces.
This ensures that long expressions can be processed by the compiler.
These optimising techniques are part of many commercial computer algebra systems, but are not part 
of the GiNaC-library. 
They have been added to this program.

\section{Proof of strategy C}
\label{appendix:proof}

In this appendix we provide a proof of strategy C. The structure of the proof is modelled
on Spivakovsky's proof of strategy B.
We keep the notations of section \ref{sect:strategies}.
\\
\\
Lemma 0:
\bq
 d\left(\Delta_k\right) & \ge & c_{k-1}.
\eq
Proof by induction: The claim is obvious for $k=0$, since $c_{-1}=0$. 
Assume that the hypothesis is true for $k$.
With $d\left(\Delta_k\right) = d\left(\tilde{\Delta}_k\right) + \left| \omega\left(\Delta_k\right) \right|$
we deduce that
$d\left(\tilde{\Delta}^C_k\right) \ge c_{k}$.
Using this it follows
$d\left(\Delta_{T_k}\right) \ge c_k$ for all $T_k$.
Therefore $d\left(\Delta_{k+1}\right) \ge c_k$.
\\
\\
Corollary: $d\left(\tilde{\Delta}_k\right) = d\left(\tilde{\Delta}_k^C\right)$.
For $c_k \ge c_{k-1}$ there is nothing to prove. For $c_k < c_{k-1}$ we have to consider the 
point $c_k \omega\left(\Delta_k\right)/(c_{k-1}-c_k)$.
From $\left|\omega\left(\Delta_k\right)\right| = d\left(\Delta_k\right) - d\left(\tilde{\Delta}_k\right) \ge c_{k-1}-c_k$
the claim follows.
\\
\\ 
We call a set $S_k$ permissible for $\Delta_k$, if
\bq
 \sum\limits_{j \in S_k} \nu_j & \ge & c_{k-1} 
 \;\;\; \mbox{for all} \; \nu \in \Delta_k.
\eq
Lemma 1: If $S_{k+1}$ is permissible for $\Delta_{k+1}$, then
$S_k = S_{k+1} \cup T_k$ is permissible for $\Delta_k$
and $d_{S_k}\left(\tilde{\Delta}_{k}\right) = d\left(\tilde{\Delta}_{k}\right)$.
To prove this lemma 
we first show
\bq
 \sum\limits_{j\in S_k} \omega_j & \ge & c_{k-1}- c_k,
\eq
where $\omega_j$ are the components of $\omega\left(\Delta_k\right)$.
For $c_k \ge c_{k-1}$ this is obvious, as the l.h.s. is non-negative.
For $c_k < c_{k-1}$ we have
\bq
 \frac{c_k}{c_{k-1}-c_k} \omega\left(\Delta_k\right) \in \tilde{\Delta}_{k}^C.
\eq
Then
\bq
 S_{k+1} \; \mbox{permissible for} \; \Delta_{k+1}
 \Rightarrow  
 \sum\limits_{j\in S_{k+1}\cup T_k} \nu_j \ge c_k \;\;\;\mbox{for all} \; \nu \in \tilde{\Delta}_{k}^C
 \;\;
 \Rightarrow \sum\limits_{j\in S_{k}} \omega_j \ge c_{k-1}-c_k.
\eq
Therefore
\bq
 S_{k+1} \; \mbox{permissible for} \; \Delta_{k+1} 
 & \Rightarrow & 
 \sum\limits_{j\in S_{k+1}\cup T_k} \nu_j \ge c_k \;\;\;\mbox{for all} \; \nu \in \tilde{\Delta}_{k}^C
 \nonumber \\
 & \Rightarrow & 
 \sum\limits_{j\in S_{k}} \nu_j \ge c_k \;\;\;\mbox{for all} \; \nu \in \tilde{\Delta}_{k}
 \nonumber \\
 & \Rightarrow &  
 \sum\limits_{j\in S_{k}} \nu_j \ge c_k + \sum\limits_{j\in S_k} \omega_j \;\;\;\mbox{for all} \; \nu \in \Delta_{k}
 \nonumber \\
 & \Rightarrow &  
 \sum\limits_{j\in S_{k}} \nu_j \ge c_{k-1} \;\;\;\mbox{for all} \; \nu \in \Delta_{k}
\eq
From the second step above
\bq
\sum\limits_{j\in S_{k}} \nu_j \ge c_k = d\left(\tilde{\Delta}_k\right) \;\;\;\mbox{for all} \; \nu \in \tilde{\Delta}_{k}
\eq
it follows immediately that
$ d_{S_k}\left(\tilde{\Delta}_{k}\right) \ge d\left(\tilde{\Delta}_k\right)$.
\\
\\
Corollary: The set $S$ as defined in section~\ref{sect:strategy_C} is permissible for $\Delta$.
For $\Delta_r \neq \emptyset$ the set $\Gamma_r$ is permissible by construction.
Then the repeated application of the above lemma shows that $S$ is permissible for $\Delta$.
For $\Delta_r = \emptyset$ it follows that $d_{S_{r-1}}\left(\tilde{\Delta}_{r-1}^C\right) \ge c_{r-1}$
and therefore $d_{S_{r-1}}\left(\Delta_{r-1}\right) \ge c_{r-1}$.
Therefore $S_{r-1}$ is permissible for $\Delta_{r-1}$ and with the same argumentation as above it follows
that $S$ is permissible for $\Delta$.
\\
\\
Let now $\Delta \subset {\mathbb R}_+^n$. We define the transformation
$ \Delta' = \sigma_{S,i}^c\left(\Delta\right)$
by
\bq
 \nu_j' & = & \nu_j \;\;\;\mbox{for} \;\; j\neq i,
 \nonumber \\
 \nu_i' & = & \sum\limits_{j\in S} \nu_j - c.
\eq
Lemma 2:
\begin{description}
\item{(a) } $d\left(\tilde{\Delta}_k'\right) \le d\left(\tilde{\Delta}_k\right)$.
\item{(b) } If $d\left(\tilde{\Delta}_k'\right) = d\left(\tilde{\Delta}_k\right)$
then $i \notin H_k$ and $H_k \subseteq H_k'$.
\item{(c) } If $d\left(\tilde{\Delta}_k'\right) = d\left(\tilde{\Delta}_k\right)$ and
$H_k=H_k'$ then $T_k=T_k'$.
\item{(d) } If $d\left(\tilde{\Delta}_k'\right) = d\left(\tilde{\Delta}_k\right)$ and
$T_k=T_k'$ then the following diagram commutes:
\begin{eqnarray*}
\begin{CD}
\Delta_k @>>> \tilde{\Delta}_k @>>> \tilde{\Delta}_k^C @>>> \Delta_{k+1} \\
@VV{\sigma_{S_k,i}^{c_{k-1}}}V @VV{\sigma_{S_k,i}^{c_k}}V @VV{\sigma_{S_k,i}^{c_k}}V @VV{\sigma_{S_{k+1},i}^{c_k}}V \\
\Delta_k' @>>> \tilde{\Delta}_k' @>>> \tilde{\Delta}_k'{}^C @>>> \Delta_{k+1}' \\
\end{CD}
\end{eqnarray*}
\end{description}
Proof: 
(a) Let $i \in S_k$.
If $i \in H_k$ there is a point $\nu \in \tilde{\Delta}_k$ such that $\left|\nu\right|=d\left(\tilde{\Delta}_k\right)$ 
and $\nu_i \neq 0$.
Then $\left|\nu'\right|< d\left(\tilde{\Delta}_k\right)$.
Assume now $i \notin H_k$ and choose a point $\nu \in \tilde{\Delta}_k$ such that $\left|\nu\right|=d\left(\tilde{\Delta}_k\right)$.
Then $\left|\nu'\right|= d\left(\tilde{\Delta}_k\right)$.
\\
(b) If $i \in H_k$ then there would be a point $\nu \in \tilde{\Delta}_k$ with
$\left|\nu'\right|< d\left(\tilde{\Delta}_k\right)$, which is in contradiction with the assumption
$d\left(\tilde{\Delta}_k'\right) = d\left(\tilde{\Delta}_k\right)$.
Therefore $i \notin H_k$.
Further for all points $\nu \in \tilde{\Delta}_k$ with $\left|\nu\right|=d\left(\tilde{\Delta}_k\right)$
we have $\left|\nu'\right|= d\left(\tilde{\Delta}_k\right)$, therefore $H_k \subseteq H_k'$.
\\
(c) is a direct consequence of (\ref{choice_T_k}).
\\
(d) is verified by a direct calculation.
\\
\\
Lemma 2 is the key to prove that the sequence (\ref{sequence_Hauser}) decreases:
According to lemma 2(a), $d\left(\tilde{\Delta}_k\right)$ either decreases or remains constants.
If it remains constant, lemma 2(b) states that $(\#I_k-\#H_k)$ either decreases or remains constant.
If also this number remains constant, lemma 2(c) guarantees $T_k=T_k'$ and lemma 2(d) allows us to descend
in dimension and to consider the simpler problem for $\Delta_{k+1}$.
\\
It is an easy exercise to complete the proof and 
to show that if $d\left(\tilde{\Delta}_r\right)=0$ the choice of $\Gamma_r$ forces
$d\left(\Delta_r\right)$ to decrease.

\end{appendix}

\bibliography{/home/stefanw/notes/biblio}
\bibliographystyle{/home/stefanw/latex-style/h-physrev3}

\end{document}